\documentclass[preprint2]{aastex}

\slugcomment{Submitted to the Astrophysical Journal}

\shorttitle{The UV SEDs of Quiescent Black Holes and Neutron Stars}
\shortauthors{Hynes \& Robinson}

\begin{document}

\title{The Ultraviolet Spectral Energy Distributions of Quiescent Black Holes and Neutron Stars}

\author{R.I. Hynes}
\affil{Department of Physics and Astronomy, Louisiana State
University, Baton Rouge, Louisiana 70803}
\email{rih@phys.lsu.edu}

\author{E. L. Robinson}
\affil{Department of Astronomy, The University of Texas at Austin, 
1 University Station C1400, Austin, Texas 78712}
\email{elr@astro.as.utexas.edu}

\begin{abstract}

  We present {\it HST}/ACS ultraviolet photometry of three quiescent
  black hole X-ray transients: X-ray Nova Muscae 1991 (GU~Mus),
  GRO~J0422+32 (V518~Per), and X-ray Nova Vel 1993 (MM~Vel), and one
  neutron star system, Aql~X-1.  These are the first quiescent UV
  detections of these objects.  All are detected at a much higher
  level than expected from their companion stars alone and are
  significant detections of the accretion flow.  Three of the four UV
  excesses can be characterized by a black body of temperature
  $5000-13,000$\,K, hotter than expected for the quiescent outer disk.
  A good fit could not be found for MM~Vel.  The source of the
  black-body-like emission is most likely a heated region of the inner
  disk.  Contrary to initial indications from spectroscopy there does
  not appear to be a systematic difference in the UV luminosity or
  spectral shape between black holes and neutron star systems.
  However combining our new data with earlier spectroscopy and
  published X-ray luminosities there is a significant difference in
  the X-ray to UV flux ratios with the neutron stars exhibiting
  $L_{\rm X}/L_{\rm UV}$ about $10\times$ higher than the black hole
  systems.  This is consistent with earlier comparisons based on
  estimating non-stellar optical light, but since both bandpasses we
  use are expected to be dominated by accretion light we present a
  cleaner comparison.  This suggests that the difference in X-ray
  luminosities cannot simply reflect differences in quiescent
  accretion rates and so the UV/X-ray ratio is a more robust
  discriminator between the black hole and neutron star populations
  than the comparison of X-ray luminosities alone.

\end{abstract}


\section{Introduction}

The discovery that many transient X-ray sources contain stellar mass
black holes has provided many opportunities to study the astrophysics
of how black holes accrete matter.  These black hole X-ray transients
(BHXRTs) and their neutron star counterparts (NSXRTs) are low-mass
X-ray binaries (LMXBs) in which a companion star, typically a dwarf or
sub-giant of solar mass or less accretes onto the compact object
through an accretion disk.  The combination of large dynamic range of
behavior observed in these objects, and accessible timescales on
which to study their variations through this range are unique to the
stellar mass objects but may also illuminate the snapshots of a particular
state seen in supermassive black holes.

One of the most interesting aspects of this study is to observe
quiescent LMXBs in which the outer parts of the accretion disk are
cool and dim and the mass transfer onto the compact object proceeds at
an extremely low rate; in XTE~J1118+480 \citet{McClintock:2003a}
inferred an accretion rate of only $10^{-8.5}$ of the Eddington
limit. Accretion in this regime is expected to proceed very
differently as the inner disc is unstable to evaporation into a
vertically extended, radiatively inefficient accretion flow.  Because
of the radiative inefficiency, it is possible for gas heated by
viscosity to either deposit the heat on the neutron star surface or
carry it through a black hole event horizon before it can radiate
away.  In the neutron star case the energy will eventually escape
whereas in the black hole case it will not, leading to the naive
expectation, apparently supported by observations, that quiescent
neutron stars should be systematically brighter than black holes
\citep{Narayan:1997a,Garcia:2001a,Hameury:2003a}.  The observational
claim has lead to great controversy, however, and objections have been
raised to this simple picture.  For example, it is necessary to
compare objects which are believed to be accreting at the same rate
from their companion stars.  \citet{Menou:1999a} argued that this was
the case if Eddington-scaled X-ray luminosities of black holes and neutron
stars of the same orbital period were compared.  Questions were also
raised about whether significant amounts of accretion energy may be
carried by the bulk motion of a jet \citep{Fender:2003a}.  Finally,
there is one NSXRT, 1H~1905+000, which appears to defy the trend
\citep{Jonker:2007a} being fainter than comparable black hole systems.
Observations beyond just a comparison of X-ray luminosities are
needed.

One promising avenue to improve on this situation is the ultraviolet.
Unlike the optical emission which appears dominated by the companion
star, UV observations with {\it HST}/FOS of a quiescent black hole
revealed a significant blue excess attributed to accretion light
\citep{McClintock:1995a}.  Subsequent more sensitive spectroscopic
observations have been performed with STIS of two black holes,
A\,0620--00 \citep{McClintock:2000a} and XTE~J1118+480
\citep{McClintock:2003a} and one neutron star, Cen~X-4
\citep{McClintock:2000a} and have confirmed this excess, as have more
recent COS observations of A\,0620--00 \citep{Froning:2011a}.
Intriguingly in the two black hole systems the flux dropped off
steeply in the UV, whereas in Cen~X-4 it actually rose (in $\nu
F_{\nu}$) leading to the suggestion that the shape of the UV spectrum
is a diagnostic of the presence of a black hole or neutron star.

Clearly such a conclusion based on a sample of three UV spectra is
unsatisfactory; but spectroscopic observations of other (fainter)
objects become prohibitively expensive, even with the Cosmic Origins
Spectrograph.  To expand the sample, we present here UV photometry
performed with {\it HST}/ACS of three more black holes, X-ray Nova Mus
1991 = GU~Mus, GRO~J0422+32 = V518~Per, and X-ray Nova Vel 1993 =
MM~Vel, and one more neutron star, Aql~X-1.  Together with the three
spectroscopic observations discussed earlier, and the UV photometry of
V404~Cyg reported separately by \citet{Hynes:2009a} this more than
doubles the sample available for comparison.

\section{Observations}
\label{ObservationSection}

Ultraviolet photometry used the Advanced Camera for Surveys (ACS;
\citealt{Gonzaga:2005a}) on the {\em Hubble Space Telescope
(HST)}.  The observations are summarized in
Table~\ref{LogTable}.  For each target 2 or 3 satellite orbits were
used, with all filters observed in each orbit in the sequence
F330W--F250W--F220W--F330W.  This provides some averaging over
variability and ensures each short wavelength observation is bracketed
by nearby F330W ones.

\placetable{LogTable}

Reduction used standard ACS techniques.  Individual images were
pre-processed to the flat-fielded stage using the automatic pipeline,
CALACS; we found no need to refine this reduction.  All of the images,
or an orbit-by-orbit subset, were then combined offline into a
geometrically corrected master image using Multidrizzle
\citep{Koekemoer:2002a} with standard settings.  We found excellent
registration between individual images.

Photometry was performed using the IDL/AstroLib aperture photometry
routine {\sc aper}.  As our targets are faint we used a relatively
small aperture to perform photometry, of radius 0.125'', with sky
defined by a 2.5'' annulus.  Other stars in the field were typically
too faint to determine per-field aperture corrections, so we use the
tabulated values from \citet{Sirianni:2005a}.  The 
aperture collects about 75\,\%\ of the total light, and almost all of
the sharp core of the point spread function.  In each case, the
position of the target was measured in the F330W bandpass and then
fixed for the other filters, after verifying that brighter stars were
consistently positioned in all the filters.

A significant source of systematic uncertainty in ACS photometry is
charge transfer inefficiency (CTI) due to radiation damage to the CCDs
(\citealt{Riess:2004a}; \citealt{Pavlovsky:2005a}).  This is worst at
low light levels, so maximized for faint sources in the UV. CTI is
also least well calibrated for these cases, and the correction
prescriptions provided by \citet{Pavlovsky:2005a} formally diverge for
negligible sky background or source counts.
For the low background case (0.5--1.5\,e$^-$/pixel) at source
brightnesses up to a few hundred electrons, \citet{Riess:2004a} found
losses of around 0.035\,mag at maximum distance from the readout
amplifier from actual data. If we rescale this to the times of our
observations and positions of the sources we find losses of 4--7\,\%. We
use these latter estimates to correct our measured source brightnesses
(or upper limits) and assign an additional error estimate of 2\,\%\
for most cases to account for the uncertainty in applying CTI
corrections. We increase this error estimate to 3\,\%\ for V518~Per as
this was the last observation made, hence subject to the largest
detector degradation, and was the faintest source. In F220W
observations that did not detect a source, the brightness is beyond
the regime calibrated by \citet{Riess:2004a} and we increase the CTI
correction error estimate to 5\,\%. The uncertainties in the CTI
corrections are always much smaller than the statistical uncertainties
in the measured fluxes and so do not significantly affect our
conclusions.

We tabulate our photometric results in Table~\ref{LogTable}.
Since we are measuring counts across a broad bandpass rather than
monochromatic fluxes, we tabulate the data in two ways: the
corrected number of electrons per second and the average flux per unit
wavelength.  Both have been corrected for CTI, and to a nominally
infinite aperture.

\section{Astrometry}

Several of our targets are known to have nearby contaminating stars in
the optical or IR which have been barely resolved. {\it HST} images
provide an ideal opportunity to obtain more precise relative
astrometry of the contaminating stars if they can be identified. We
detected the stars near MM~Vel \citep{Filippenko:1999a} and Aql~X-1
\citep{Chevalier:1999a}. We were unable to detect the star north-east
of V518~Per reported by \citet{Reynolds:2007a}.

\begin{figure*}
\begin{center}
\includegraphics[scale=0.5]{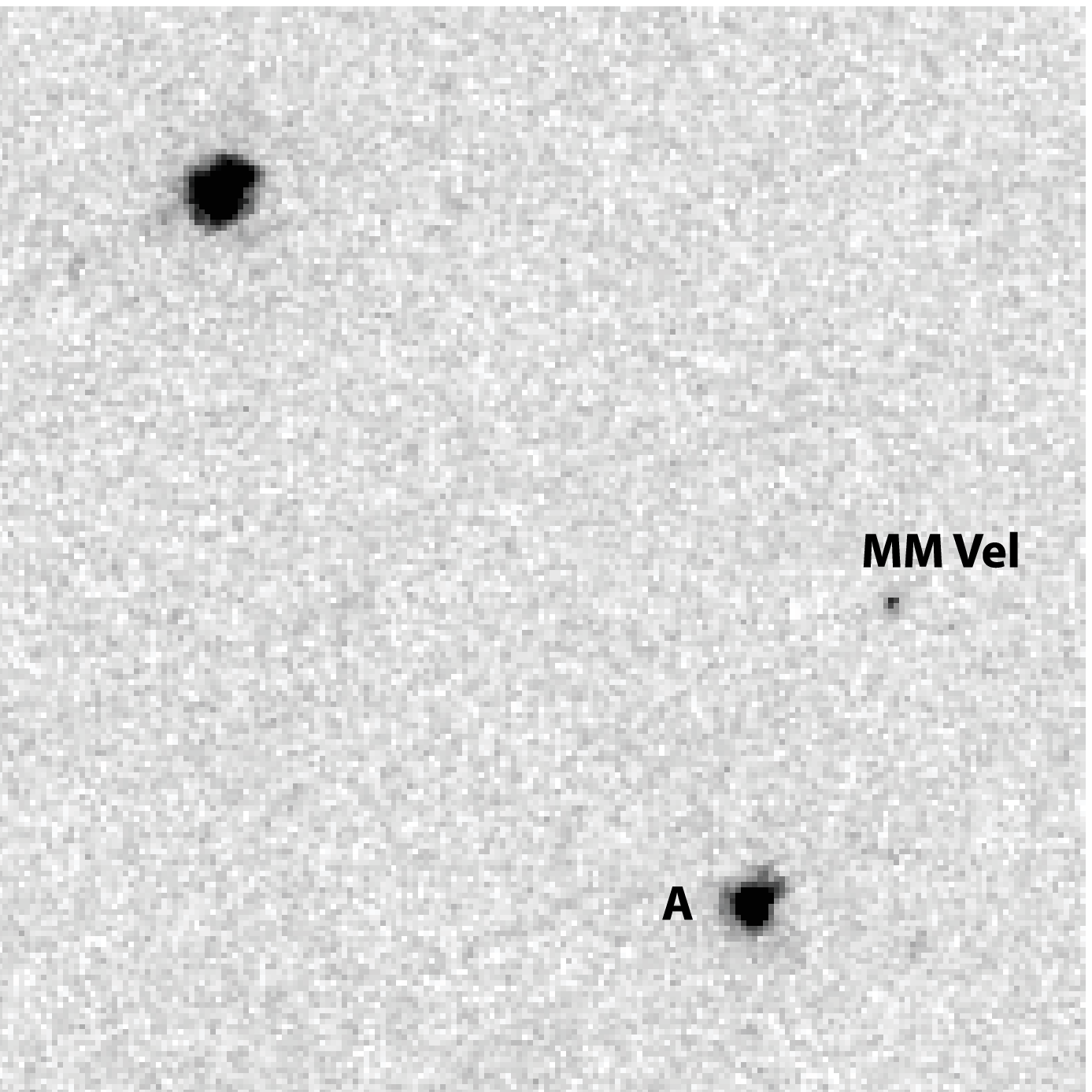}\hspace*{\fill}
\includegraphics[scale=0.5]{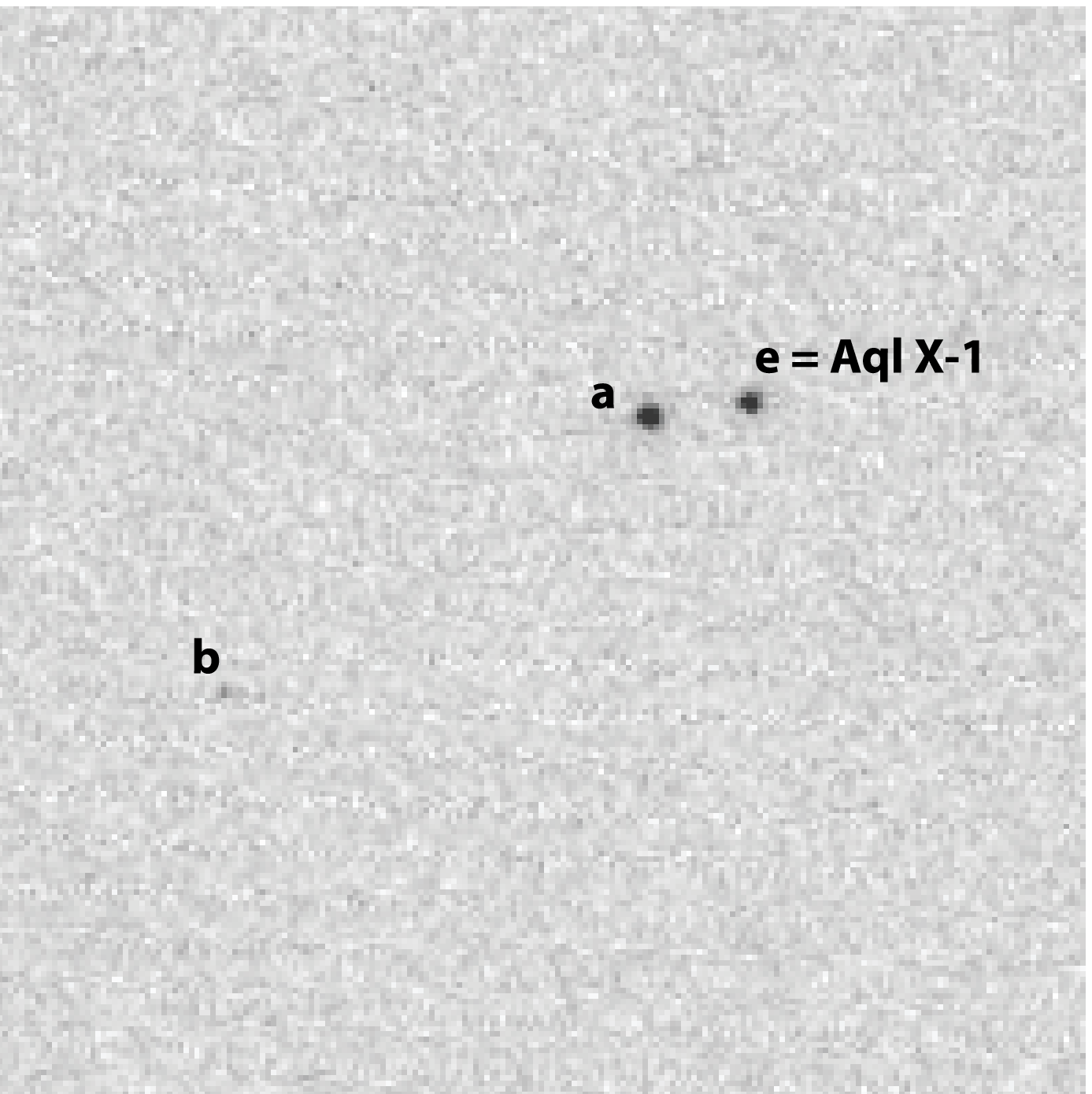}
\end{center}
\caption{F330W images of the field of MM~Vel (left) and Aql X-1 (right).  The fields are 5''
  square, and labels are from \citet{Filippenko:1999a} and \citet{Chevalier:1999a}.  The
  orientation is conventional (north at top and east to the left).  This means that the image of Aql~X-1 
is inverted with respect to that of \citet{Chevalier:1999a}.}
\label{FinderFig}
\end{figure*}

The F330W image of MM~Vel is shown in Fig.~\ref{FinderFig}.  Star
identifications are taken from \citet{Filippenko:1999a}.  The close
blend of stars seen by \citet{Filippenko:1999a} is well resolved and
no new stars are detected.  The separation of MM~Vel from star A is
$\Delta\alpha=-0.048$\,s, $\Delta\delta=+1.37$'', corresponding to a
separation 1.47'', roughly consistent with the estimate of
\citet{Filippenko:1999a} that MM~Vel was 1.6\,arcsec north-north-west
of star A.

\placefigure{FinderFig}

We show the F330W image of Aql~X-1 in Fig.~\ref{FinderFig}.  Star
identifications are taken from \citet{Chevalier:1999a}.  Aql~X-1 is
star (e), much brighter relative to star (a) than it is in the $I$ band.
Star (b) is weakly detected, but stars (c) and (d) are undetected in this
bandpass.  We find no new stars in the immediate vicinity of Aql~X-1.
The separation of Aql~X-1 from star (a) is $\Delta\alpha=-0.031$\,s,
$\Delta\delta=+0.07$'', corresponding to a separation 0.48'',
consistent with that found by \citet{Chevalier:1999a}.

\section{Adopted Parameters and Archival Measurements}
\label{ParameterSection} 

\subsection{GU~Mus = X-ray Nova Muscae 1991}

For the reddening and distance to GU~Mus, we follow
\citet{Hynes:2005a} and do not repeat the arguments made there; see
that work for the original references.  These were a reddening of
$E(B-V)=0.30\pm0.06$ and distance $d=5.89\pm0.26$\,kpc.  The spectral
type of GU~Mus has been estimated as K3--K5V \citep{Orosz:1996a} and
K3--K4V \citep{Casares:1997a}.  A K4V classification implies an
effective temperature around 4550\,K interpolating on the tables in
\citet{Cox:2000a}.  Using the system parameters from
\citet{Gelino:2001a}, including a 10.38\,hr orbital period, we expect
a surface gravity of $\log g=4.32$.

Quiescent photometry was compiled by \citet{Gelino:2001a} from both their
own observations and those of \citet{Remillard:1992a}, \citet{Orosz:1996a},
\citet{King:1996a}, and \citet{dellaValle:1998a}.  These results are
quite varied, with $V$ magnitudes ranging from 20.35 to 20.83.  This
variability is consistent with the large amplitude quiescent
variations seen by \citet{Hynes:2003a} and indicates that
non-simultaneous optical photometry cannot reliably be combined with
our UV data.

\subsection{V518~Per = GRO~J0422+32}

As for GU~Mus, we follow \citet{Hynes:2005a} and adopt
$E(B-V)=0.35\pm0.10$ and $d=2.49\pm0.30$\,kpc.  Opinion on the
spectral type of V518~Per is divided.  Early estimates favored early M
spectral types: M0--M4V \citep{Casares:1995a} or M1--M4V
\citep{Harlaftis:1999a}.  \citet{Webb:2000a} favored slightly later
types, M4--M5V, but \citet{Gelino:2003a} argued that the spectral
energy distribution was best fit by an M1V star.  \citet{Gelino:2003a}
assumed negligible disk contamination even in the optical, however,
whereas the other authors cited found significant amounts (above
50\,\%), suggesting that a later spectral type with blue contamination
may indeed be a better description.  In fact, \citet{Reynolds:2007a}
find substantial flickering even in the near-IR, so a spectral type
determined from the SED should be viewed with considerable caution.
We therefore consider both M1V and M5V classifications, adopting
effective temperatures of 3680\,K and 3170\,K respectively.  Using the
system parameters of \citet{Gelino:2003a}, including a 5.09\,hr
orbital period, implies a surface gravity of $\log g=4.66$.

For optical/IR photometry we use data from \citet{Zhao:1994a},
\citet{Casares:1995a}, \citet{Garcia:1996a}, \citet{Callanan:1996a},
\citet{Chevalier:1996a}, \citet{Gelino:2003a}, and
\citet{Reynolds:2007a}.  As for GU~Mus, substantial variations occur
from epoch to epoch.

\subsection{MM~Vel = X-ray Nova Velorum 1993}

For MM~Vel, we follow \citet{Hynes:2005a} and adopt
$E(B-V)=0.20\pm0.05$ and $d=3.82\pm0.27$\,kpc. The most persuasive
spectral type determination for MM~Vel is that by
\citet{Filippenko:1999a} who find a spectral type of K7--M0V, although
not rejecting K6V strongly. Other authors preferred earlier spectral
types \citep{Shahbaz:1996a,dellaValle:1997a}. We will adopt K7V as a
compromise, implying an effective temperature of 4180\,K. Using the
system parameters of \citet{Gelino:2004a} together with the 6.86\,hr
orbital period of \citet{Shahbaz:1996a} implies a surface gravity of
$\log g=4.55$.

Published quiescent photometry of this object is much sparser than for
GU~Mus and V518~Per, limited to $R$ band measurements of
\citet{Shahbaz:1996a} and \citet{Filippenko:1999a}, and $V$ band by
\citet{Hynes:2003a}.

\subsection{Aql X-1}

The interstellar reddening to Aql~X-1 has been estimated at
$E(B-V)=0.5\pm0.1$.  Although \citet{Chevalier:1999a} estimate a
distance of $d=2.5$\,kpc, \citet{Rutledge:2001a} argue that this is
too low and instead obtain a distance of 4.0--6.5\,kpc by requiring
that the companion must fill its Roche lobe, and using the peak
luminosity of photospheric radius expansion X-ray bursts.  They adopt
a preferred distance of 5\,kpc and we follow this.  The companion star
was estimated by \citet{Chevalier:1999a} to be a K7V star.  With an
18.95\,hr period, 1.4\,M$_{\odot}$ neutron star, and mass ratio
$q=0.33$ \citep{Welsh:2000a} we expect $\log g=3.93$.  Plausible
uncertainties in $q$ have a very small effect on this.

Like MM~Vel, Aql~X-1 is very crowded in ground-based observations
making optical photometry rare and somewhat uncertain.  The primary
source is \citet{Chevalier:1999a}.

\section{Spectral Energy Distributions}

\subsection{Fitting Methodology}

We begin by fixing the spectral type and reddening to the values
estimated in Section~\ref{ParameterSection}.  We then adopt a suitable
spectrum from \citet{Hauschildt:1999a} for the companion star, redden
it using the extinction law of \citet{Fitzpatrick:1999a}, and evaluate
synthetic photometry of the reddened companion in the bandpasses of
interest using the {\sc synphot} synthetic photometry package
\citep{Laidler:2005a}.  One of the primary advantages of this approach
is to correctly allow for red leak from the companion star's red light
into the UV bandpasses.  In a similar way, we calculate models for the
accretion light over the range of parameters of interest, redden them,
and perform synthetic photometry.  This gives us tabulated fluxes for
the two components, allowing us to fit the composite model to the UV
photometric data in its native form with the normalization of each
component as a free parameter.

We adopt black body models for the disk emission. A single-temperature
model provides the simplest characterization of the data.  For
comparison with the results of \citet{McClintock:2003a} on
XTE~J1118+480, we also fit models where the UV emission comes from the
inner edge of a multicolor black body disk.  Note that this model
explicitly assumes that the accretion rate is independent of radius
(steady-state), so this may not be an accurate description of the
behavior of a quiescent disk.  In the case of GU~Mus, we also compare
a model of a slab of recombining hydrogen implemented in {\sc
  synspec}, described by an effective temperature, a column density,
and an area.  It can be seen from
Figs.~\ref{NMusFitFig}--\ref{AqlFitFig} that although we have
considered the contribution of the companion star, both directly and
via its red-leak, in every case the companion contributions in the UV
are much smaller than those from the accretion disk and so have very
small impact on our fitting results.

\subsection{GU~Mus = X-ray Nova Muscae 1991}

Since the spectral type of the secondary star in GU~Mus is relatively
well constrained we fix its spectral type. Published optical
photometry spans rather a large range, but this is consistent with
observed short term variability \citep{Hynes:2003a}. We
fix the brightness of the companion star spectrum to lie at the
lower envelope of the optical points. The brightness adopted is not
critical, as this results in only a 5\,\%\ contribution to the
F330W filter, and negligible impact in other UV passbands.

\begin{figure*}
\begin{center}
\includegraphics[angle=90,scale=0.4]{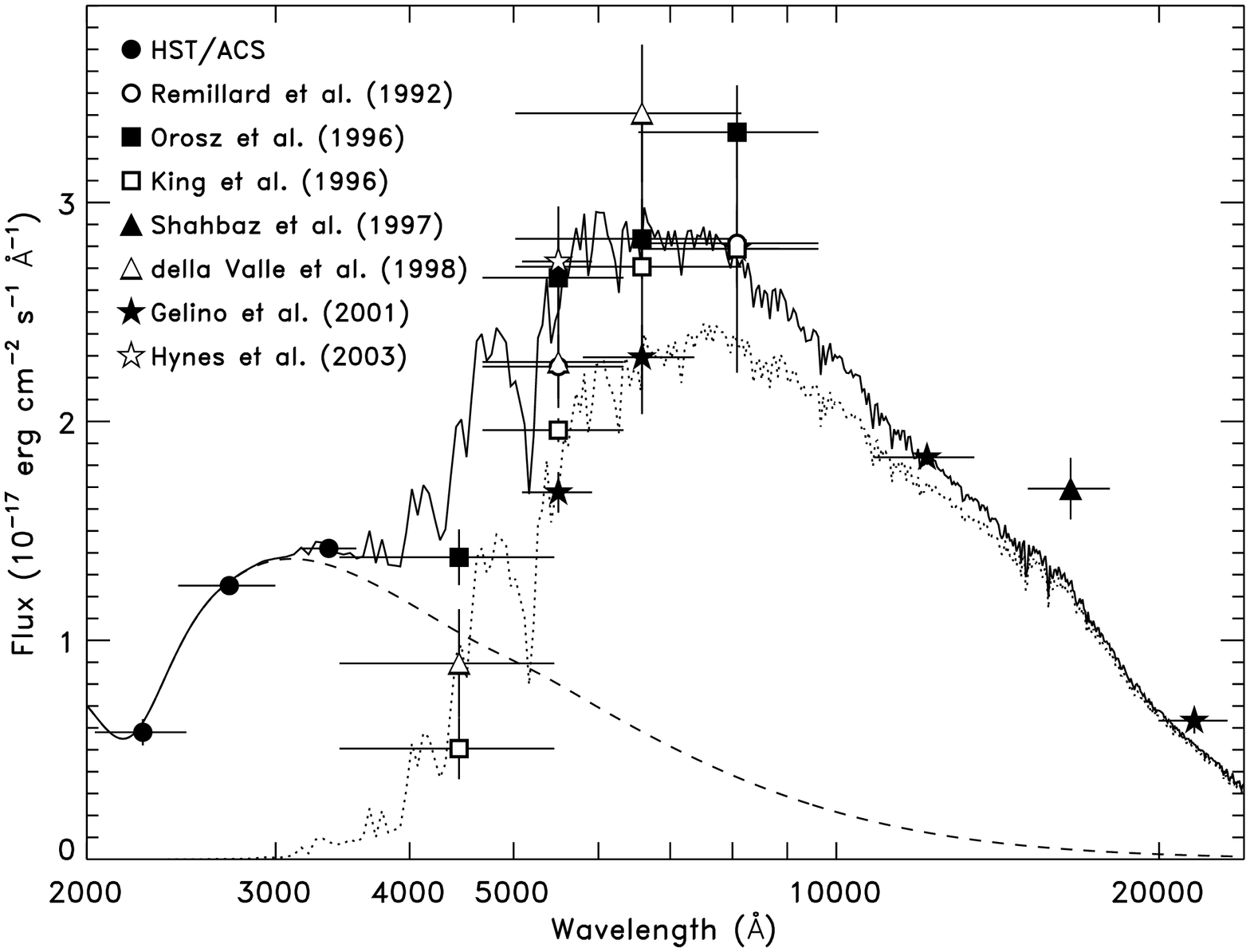}\\
\includegraphics[angle=90,scale=0.4]{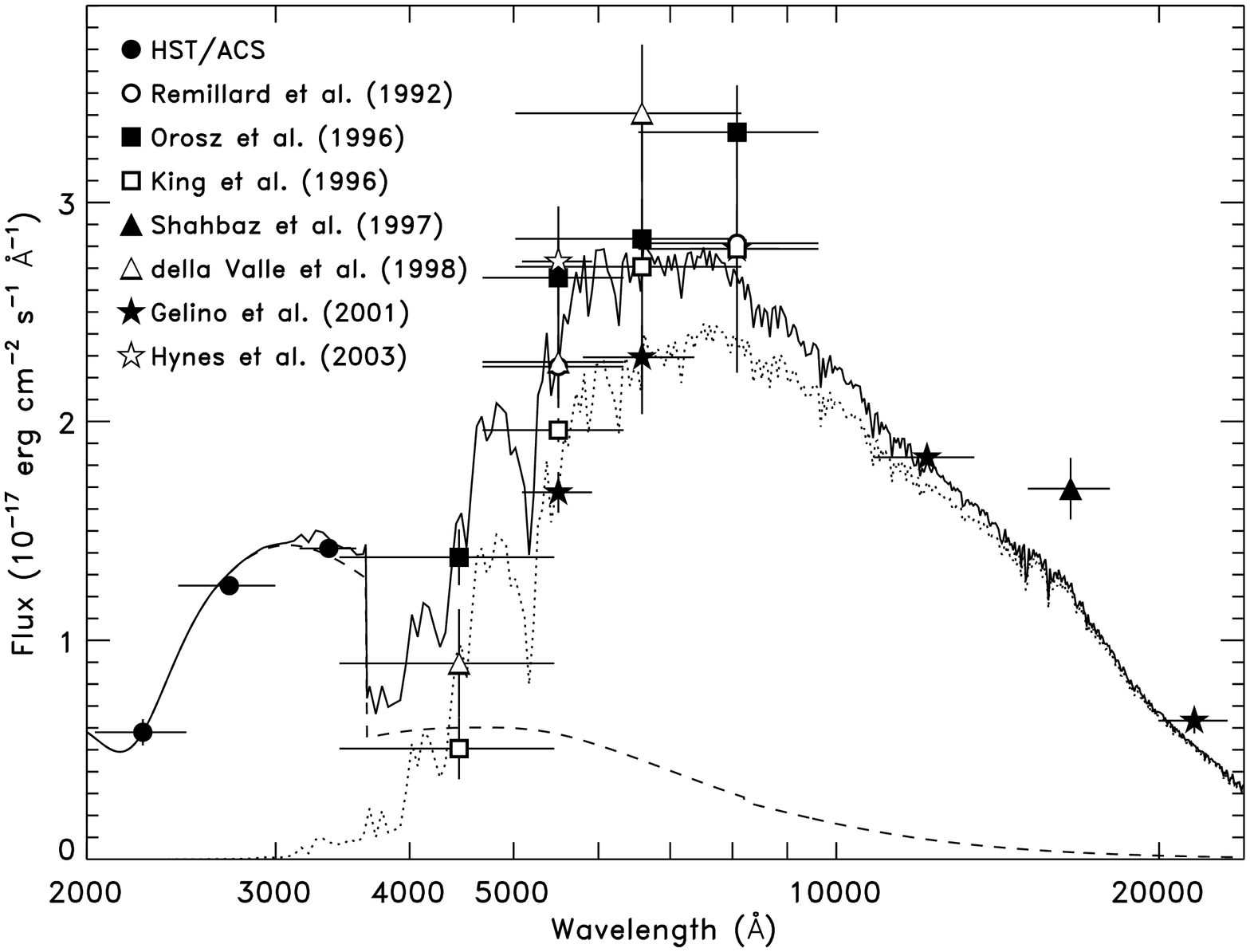}
\end{center}
\caption{Black body plus stellar fit to data on GU~Mus (upper panel)
  and optically thin recombination spectrum plus stellar fit (lower
  panel).  Wavelength uncertainties shown on the photometric points
  represent the FWHM of the filter bandpasses.  In this and subsequent
  plots, the dotted line represents the stellar component, the dashed
  line is the accretion component, and the solid line is the sum of
  the two.}
\label{NMusFitFig}
\end{figure*}

The residual photometric SED visually appears well described by a
single-temperature blackbody with temperature $T=13000\pm1400$\,K and
area about 0.3\,\%\ of the projected area of a tidally truncated
accretion disk ($\sim0.9R_{\rm Roche}$; \citealt{Whitehurst:1991a}) at
5.89\,kpc. This fit is shown in Fig.~\ref{NMusFitFig}.  Formally the
fit yields $\chi^2=8.25$ with one degree of freedom, so the fit is not
as good as might be expected. The inferred blackbody is small compared
to the accretion disk, and a region of radius 6\,\%\ of the maximal
accretion disk radius could readily account for this. It could be
interpreted as either the accretion stream-impact point, or a hot
inner annulus of the disk. The effect of uncertainty in the companion
star is indeed small. Reducing its contribution to zero, or doubling
it, changes the derived temperature by less than 1000\,K.

\placefigure{NMusFitFig}

For comparison to XTE~J1118+480, we also performed an analogous fit with
a multi-color black body disk.  The best fit was obtained with $T_{\rm
  in}=21500\pm5700$\,K, with the observed flux correspoding to $R_{\rm
  in} \simeq 3\times10^9$\,cm.  This is hotter than found by
\citet{McClintock:2003a}, but the inner disk radius is actually
inferred to be about the same.

Since the inferred area is so low, we can also consider the
possibility that the light comes from a larger but optically thinner
region of the disk. We should be cautious as we have introduced an
additional free parameter and are now fitting three photometric bands
with a three-parameter model. With this caveat, we find a
statistically better fit with the optically thin model than with the
black body, $\chi^2=3.34$ with no degrees of freedom. The best fit is
found for a column density of $\sim10^{20}$\,cm$^{-2}$ and temperature
$T=16,100\pm2000$\,K. Because the temperature is higher, while the
optical depth remains relatively high, the required surface area is
actually a little smaller than for the blackbody fit, corresponding to
a region of radius about 4\,\%\ of the maximal disk radius. Thus
whether the UV is modeled by an optically thick or thin component, we
come to the same conclusion: the emission originates from a hot region
with a rather small projected cross-sectional area compared to that of
the disk.

A potential advantage of the optically thin model can be seen in
Fig.~\ref{NMusFitFig}. The blackbody model predicts a substantial disk
contribution in $B$, much higher than observed by several groups. The
optically thin model predicts a much lower $B$ flux than the blackbody
model, so is much easier to reconcile with the archival optical
photometry. Since the optical photometry were not simultaneous with
the UV data, however, an alternative explanation is that the UV
observations occurred when the accretion light was brighter than seen
in the optical photometry. This is a quite plausible explanation,
since the source is known to exhibit large amplitude optical flaring
\citep{Hynes:2003a}. As discussed in Section~\ref{LuminositySection}
and shown in Fig.~\ref{LuminosityFig}, GU~Mus also has the highest
inferred UV luminosity of the sample here, consistent with this being
an unusually high state.

An additional argument against attributing the UV emission of
quiescent BHXRTs to optically thin emission (in general, rather than
in the specific case of GU~Mus) is that this model predicts a large
Balmer jump in emission. While our photometric observations cannot
discriminate this, spectroscopic observations of other sources have
been performed and reveal no such Balmer continuum emission. In both
A\,0620--00 \citep{McClintock:1995a} and XTE~J1118+480
\citep{McClintock:2003a} simultaneous data either through or on either
side of the Balmer jump shows no significant discontinuity, supporting
an optically thick interpretation at least in these cases. In view of
this evidence, we believe that the optically thick (black body) model
is the most credible and apply this to the remaining sources. It is
certainly possible, of course, that the optical depth varies from
source to source, but without strictly simultaneous coverage across
the Balmer jump we cannot constrain this.

\subsection{V518~Per = GRO~J0422+32}

\begin{figure*}
\begin{center}
\includegraphics[angle=90,scale=0.4]{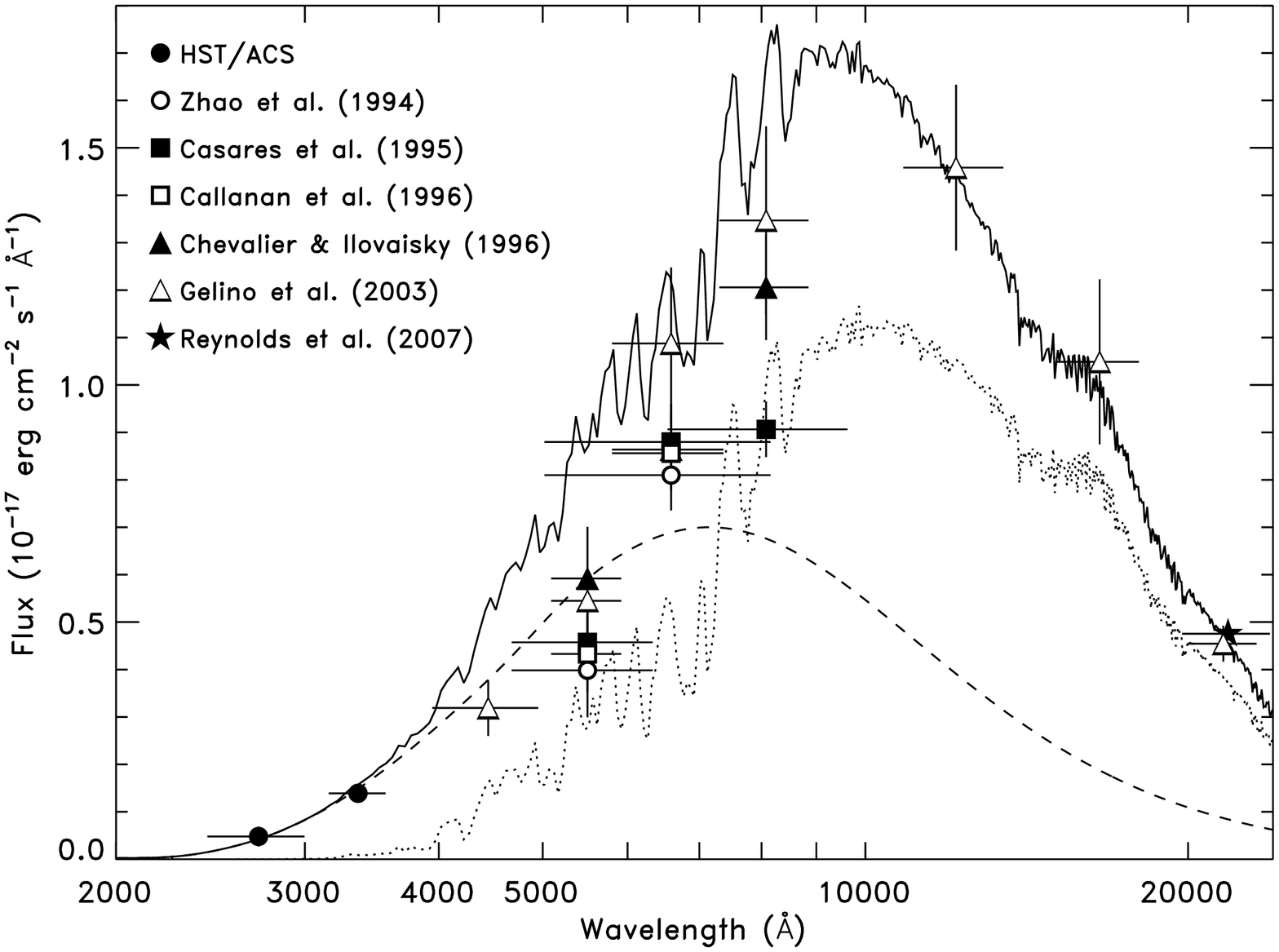}\\
\includegraphics[angle=90,scale=0.4]{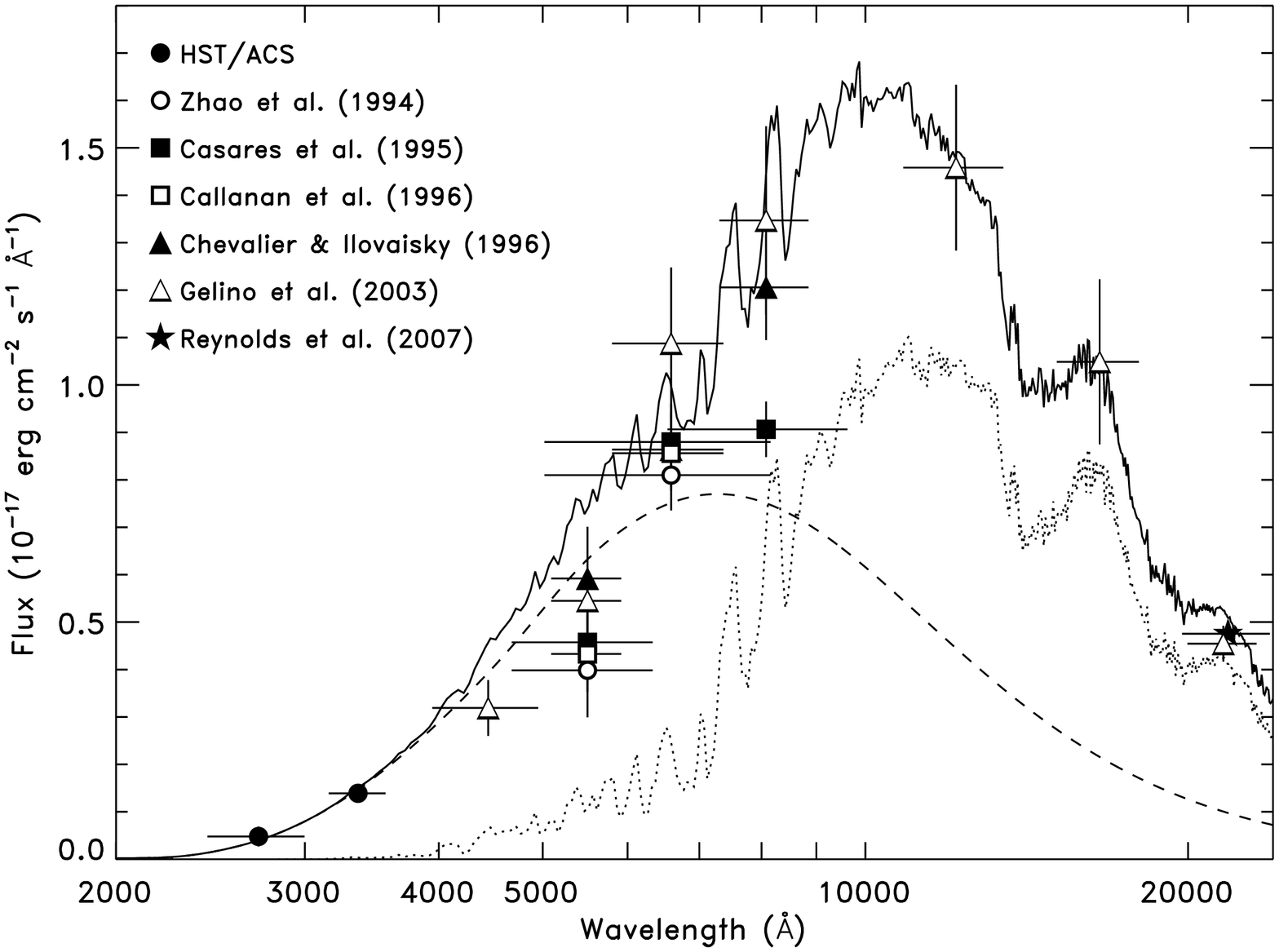}
\end{center}
\caption{Black body plus stellar fit to data on V518~Per.  The upper
  panel is for an M1V companion, the lower panel for M5V.}
\label{NPerFitFig}
\end{figure*}

We consider both M1V and M5V spectral types for V518~Per
(Fig.~\ref{NPerFitFig}). Allowing that the strength (and possibly
temperature) of the hot component could vary, both possibilities are
acceptable, although M5V appears to provide a better fit in the
optical region of the SED.  Both models suggest a large or dominant
accretion contribution in the optical, and a measurable one in the
infrared, in agreement with observations.

\placefigure{NPerFitFig}

Both spectral types produce negligible UV light (even allowing for red
leaks) and consequently, the hot component derived is largely
independent of the assumed companion spectral type. For an M1
companion we find $T=5200\pm2200$\,K, whereas for M5 we find
$T=5100\pm2000$\,K. In both cases, the hot component has an area
0.3\,\%\ of the projected area of a tidally truncated disk at
2.49\,kpc (as was also found in GU~Mus).  If we instead use a
multi-color black body disk model, we obtain a similar temperature,
$T_{\rm in}=5800\pm2900$\,K, with a rather larger inner radius than
GU~Mus, $R_{\rm in}\simeq 13\times10^9$\,cm.

\subsection{MM~Vel = X-ray Nova Velorum 1993}

\begin{figure*}
\begin{center}
\includegraphics[angle=90,scale=0.4]{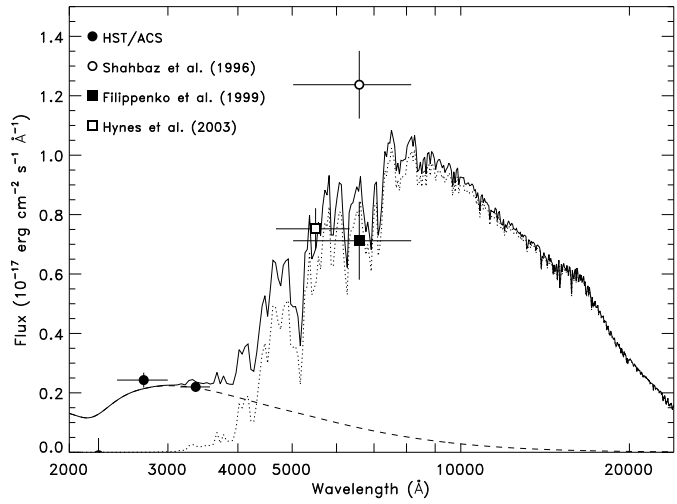}
\end{center}
\caption{Black body plus stellar fit to data on MM~Vel}
\label{NVelFitFig}
\end{figure*}

While a UV excess is strongly detected in MM~Vel, we encountered
difficulties in properly fitting it.  Neither the black body nor
optically thin recombination model could reproduce the combination of
a rising spectrum from F330W to F250W and the non-detection in F220W.
We show in Fig.~\ref{NVelFitFig} a representative 12,000\,K model but
do not feel it appropriate to quote a formal fit in this case.  This
model over-predicts the F220W flux, but since the discrepancy is less
than 3\,$\sigma$ this may be a statistical fluctuation.  We note that
the source size implied by a 12\,000\,K black body model is smaller
than in the other systems in this paper, just 0.06\,\%\ of the
estimated projected area of a tidally truncated disk, which may be
another indication that there are differences in the source of the UV
emission in this system.  Using a multi-color black body disk model we
derive rather an inner temperature of $T=17800$\,K, with a lower limit
($1-\sigma$) of 9100\,K, and the upper limit unconstrained.  For the
best fit, the inner disk radius derived is $R_{\rm in}\simeq
1.1\times10^9$\,cm.

\placefigure{NVelFitFig}

It is possible that the F250W flux is high due to the contribution
from the Mg\,{\sc ii} line which is known to be strong in quiescent
LMXBs \citep{McClintock:1995a,McClintock:2000a,McClintock:2003a}.
Another explanation for the discrepancy is intrinsic source
variability. To test this, we examined individual F330W observations
(which by design bracketed the F250W and F220W observations). We found
the F330W flux showed a standard deviation of only 15\,\%\ of the
mean with no systematic trend, small enough to be a constant flux
within uncertainties. This does not rule out variability as an
explanation, either for a low F220W flux or a high F250W one, but
provides no evidence for such an explanation.

\subsection{Aql X-1}

\begin{figure*}
\begin{center}
\includegraphics[angle=90,scale=0.4]{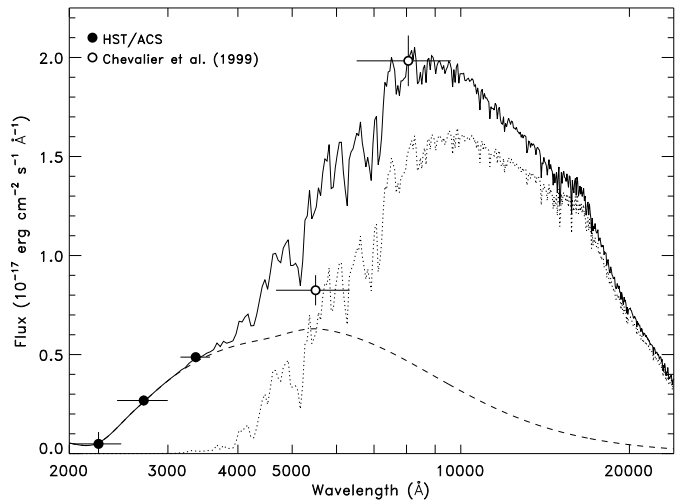}
\end{center}
\caption{Black body plus stellar fit to data on Aql~X-1}
\label{AqlFitFig}
\end{figure*}

Aql~X-1 has the least well-constrained optical SED of the sources
considered, but as in the other cases, this has relatively little
impact on the interpretation of the UV data as all plausible companion
star spectra make negligible UV contribution. We can obtain a good fit
to the UV data with a black body of $T=9300\pm700$\,K and about
7.5\,\%\ of the area of a tidally truncated disk.  The available data
and model fit are shown in Fig.~\ref{AqlFitFig}.  With the alternative
multi-color black body fit, we obtain $T_{\rm in}=12400\pm1500$\,K and
$R_{\rm in}\simeq 7\times10^9$\,cm.

\placefigure{AqlFitFig}

\section{Discussion}

\subsection{Comparison between UV and X-ray luminosities}
\label{LuminositySection}

\begin{figure*}
\begin{center}
\includegraphics[angle=90,scale=0.6]{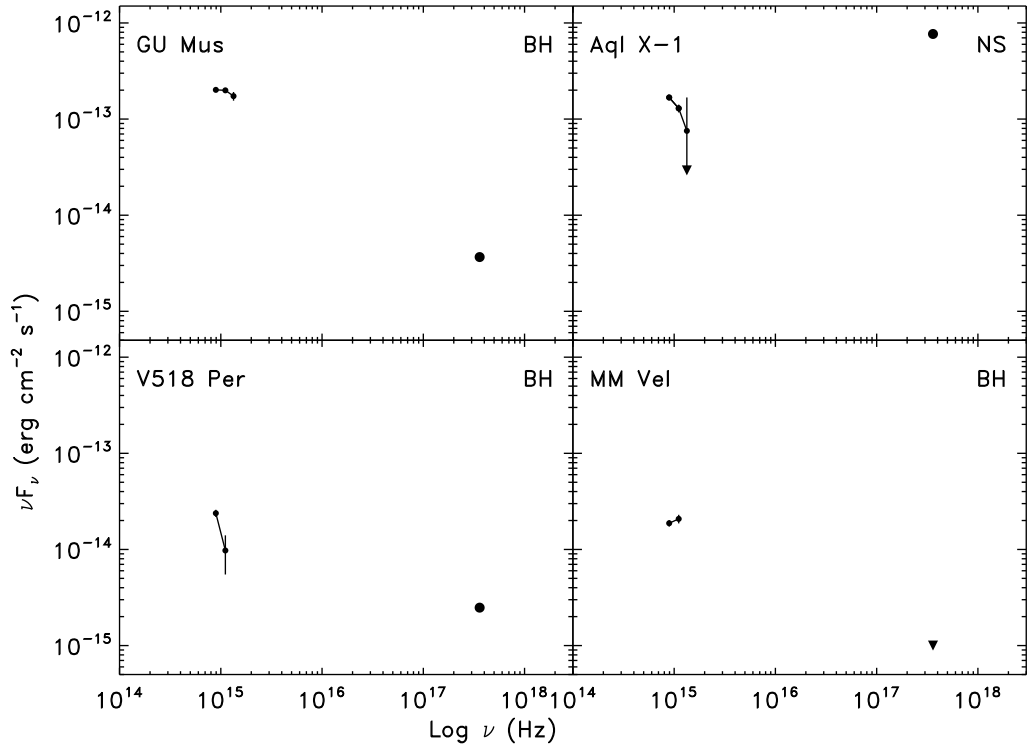}
\end{center}
\caption{UV--X-ray spectral energy distributions of the four targets
  observed here.  We do not show the F220W non-detections of V518~Per
  and MM~Vel.  We also include X-ray measurements taken from \citet{Sutaria:2002a} for GU~Mus,
\citet{Garcia:2001a} for V518~Per, \citet{Hameury:2003a} for MM~Vel, and
\citet{Narayan:1997a} for Aql~X-1.   The black holes all decline in $\nu F_{\nu}$ from UV
to X-rays, the one neutron star increases.}
\label{AllSEDFig}
\end{figure*}

We compile the SEDs in $\nu F_{\nu}$ form in Fig.~\ref{AllSEDFig}.
\citet{McClintock:2003a} suggested that the shape of the UV spectrum
could discriminate between black hole and neutron star systems.  That
does not seem to be borne out by our larger sample where it instead
appears that the shape of the UV spectrum reflects random variance in
the location of the peak of the spectrum from system to system.  Note
in particular that both GU~Mus and Aql~X-1 can be well fitted by black
body models, but that it is the black hole system, not the neutron
star, which has the higher temperature of the pair.  As can be seen
from Fig.~\ref{AllSEDFig}, however, our sample does seem to bear out
the trend that the X-ray to UV ratio is higher in NSXRTs than in their
black hole counterparts.  Both Cen~X-4 \citep{McClintock:2000a} and
Aql~X-1 show SEDs rising from UV to X-ray, whereas the black hole
systems all decline.  This result in itself is not new, and bears out
the earlier realization that black holes have optical/UV non-stellar
luminosities exceeding their X-ray luminosities, whereas in neutron
stars this is reversed (\citealt{Campana:2000a}; see also
\citealt{Narayan:2008a}).  The advantage we have, however, is that by
working with the almost pure UV accretion light, we are not at the
mercy of uncertainties in the stellar contribution; we can directly
compare UV accretion light with X-ray emission.

\placefigure{AllSEDFig}

\begin{figure*}
\begin{center}
\includegraphics[angle=90,scale=0.4]{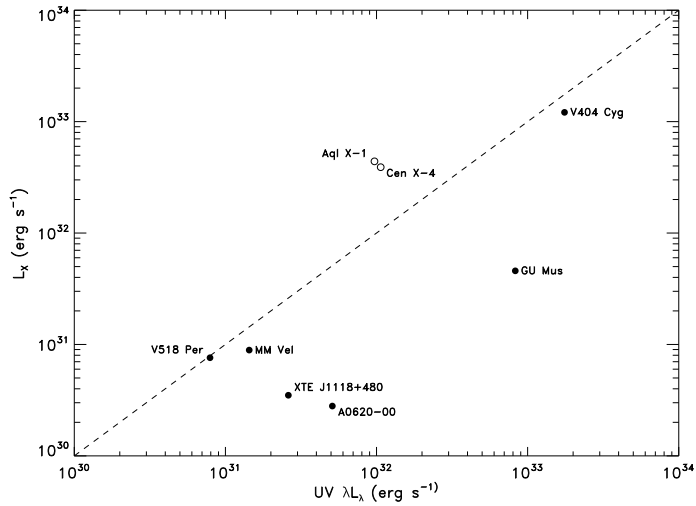}
\end{center}
\caption{The relationship between UV and X-ray luminosities of six
  quiescent BHXRTs (filled circles) and two NSXRTs (open circles).
  The dashed line corresponds to $\lambda L_{\lambda}=L_{\rm X}$.
  Lines parallel to this are loci of constant $L_{\rm X} / L_{\rm
    UV}$.  Uncertainties in distance will also move a source parallel
  to this line.  Uncertainties in dust extinction will move a source
  horizontally in the diagram.}
\label{LuminosityFig}
\end{figure*}

We can perform the comparison more rigorously by examining the
relationship between X-ray and (dereddened) UV luminosities. In
Fig.~\ref{LuminosityFig} we compile these data for the four sources in
our sample, and for archival STIS spectra of A\,0620--00, Cen~X-4, and
XTE~J1118+480 \citep{McClintock:2000a,McClintock:2003a}, and ACS
photometry of V404~Cyg \citep{Hynes:2009a}. X-ray data are taken from
\citet{Garcia:2001a} for A\,0620--00 and V518~Per, from
\citet{Campana:2004a} for Cen~X-4, from \citet{Narayan:1997a} for
Aql~X-1, from \citet{Sutaria:2002a} for GU~Mus, from
\citet{Hameury:2003a} for MM~Vel, from \citet{McClintock:2003a} for
XTE~J1118+480 and from \citet{Hynes:2009a} for V404~Cyg. It should be
noted that only XTE~J1118+480, and V404~Cyg have {\em simultaneous} UV
and X-ray measurements. As anticipated, the neutron star systems
appear systematically above those containing black holes (ranks 1 and
2 out of 8 ranked by $L_{\rm X} / L_{\rm UV}$). All neutron stars have
$L_{\rm X} > L_{\rm UV}$ and all black holes have $L_{\rm X} < L_{\rm
  UV}$.  Note in particular that GU~Mus is securely detected at a much
higher UV luminosity than either of the NS systems, even though it is
much fainter at X-ray energies.

\placefigure{LuminosityFig}

This suggests that black hole systems are less efficient at producing
X-rays from accreted material than neutron stars (assuming that the UV
luminosity is a measure of the accretion rate feeding the compact
objects). This is not a new idea, of course, and similar claims have
been made based on a comparison of X-ray luminosities alone or in
combination with estimates of the non-stellar optical light
\citep{Narayan:1997a,Campana:2000a,Garcia:2001a,Hameury:2003a,Narayan:2008a}.
However, comparison of the UV flux with the X-ray flux gives us more
confidence that differences reflect efficiency of production of
X-rays, rather than differences in mass accretion rate.  Furthermore,
in the case of Aql~X-1, the UV properties are very similar to those of
the black hole sample, suggesting that the difference between the two
is indeed driven by X-ray differences rather than UV ones.

\subsection{The UV light source}

Where we are able to obtain good fits to the SEDs, they are
characterized by temperatures higher than expected in quiescent
accretion disks, which should be $\la 3000$\,K \citep{Menou:2002a},
and have much smaller emitting areas than the accretion disk.
Previously published spectroscopic observations show similar trends.
A\,0620--00 was fitted by \citet{McClintock:1995a} with a 9,000\,K
black body with area $1/80$th of that of the accretion disk.
XTE~J1118+480 was described by \citet{McClintock:2003a} as being
fitted with a multicolor black body disk model with inner edge
temperature 13,000\,K. Cen~X-4 exhibits a UV spectrum that rises
monotonically in $\nu F_{\nu}$, indicating a hot source of emission
\citep{McClintock:2000a}. The most likely nature of a localized, hot
region of UV emission would seem to be the accretion stream impact
point, but a hot inner edge to the accretion disk near a transition to
an evaporated flow might be another possibility.

If the UV emission originates at the stream-impact point we would
expect the UV luminosity to come from energy liberated in falling to
the stream-impact point.  For the luminosity of the hot source we
integrate the unreddened black body fit to the UV photometry.  We can
estimate the maximum energy we might expect to release as half the
binding energy of material at the circularization radius in the
accretion disk. If the disk extends beyond the circularization radius,
then the stream-impact point will be higher in the potential well and
less energy will be liberated.  For GU~Mus, we require a very high
accretion rate of about $4\times10^{-9}$\,M$_{\odot}$\,yr$^{-1}$ to
explain the UV luminosity if
this is sustained. This can be compared to the estimated mass required
to power the 1991 outburst of $10^{-8}$\,M$_{\odot}$. At this
accretion rate, the mass required by the outburst could be supplied in
$\sim3$ years, whereas we have only seen one outburst of GU~Mus,
suggesting a recurrence time at least ten times larger than this. If
the energy source of the UV emission is to be gravitational energy
liberated at the hot-spot, we then require that the accretion rate,
and UV luminosity, be highly variable and that the GU~Mus observations
were well above the average level as suggested earlier.

Performing similar calculations for the other sources yields mass
transfer rates from the donor star of
$3\times10^{-10}$\,M$_{\odot}$\,yr$^{-1}$ for V518~Per,
$1\times10^{-10}$\,M$_{\odot}$\,yr$^{-1}$ for MM~Vel, and
$3\times10^{-9}$\,M$_{\odot}$\,yr$^{-1}$ for Aql~X-1.  The values for
V518~Per and MM~Vel are substantially below that of GU~Mus, and do not
impose strong constraints, but that for Aql~X-1 is comparably high.
In this case, we have observed multiple outbursts, and so have a much
more secure estimate of the time-average accretion rate.
\citet{Rutledge:2001a} have estimated this at
$2.5\times10^{-10}$\,M$_{\odot}$\,yr$^{-1}$ (assuming a 5\,kpc distance
as we have used).  This again is a discrepancy of an order of
magnitude with the observed UV brightness.

Both GU~Mus and Aql~X-1 would require mass transfer rates from their
donor stars in excess of $10\times$ the plausible average rate to
explain the UV light source as a stream-impact point at or outside the
circularization radius in the disk, and so this explanation seems
untenable.  An alternative explanation is that the UV originates much
deeper in the gravitational potential well, in the inner disk just
outside the transition radius, as suggested by \citet{Campana:2000a}
and specifically considered for XTE~J1118+480 by
\citet{McClintock:2003a}.  Fitting our photometry with multi-color
black body disk models yields comparable results to XTE~J1118+480,
with high temperatures and radii of order 1--10$\times10^9$\,cm,
comparable to expected transition radii of 1000-10,000\,$R_{\rm sch}$.
In this case a lower mass accretion rate is needed to supply the
energy than if the UV emission originates at the stream-impact
point. If some of the accretion stream overflows the disk, it will
impact closer to the compact object and could provide an alternative
mechanism for increased heating of the inner disk although in this
case we might expect to see evidence for stream-overflow in
phase-resolved spectroscopy of quiescent LMXBs.  We note that another
explanation, that the inner disk is simply heated by the X-ray source,
cannot adequately explain the UV emission when we see UV luminosities
comparable to or exceeding the X-ray luminosity.


\section{Conclusions}

We have found UV excesses in four quiescent LMXB systems, three black
hole systems and one containing a neutron star. In every case, the UV
detection is secure and greatly exceeds that expected by the companion
star. The spectral shapes are heterogeneous, and black body models
require a variety of temperatures and emitting areas. In some cases a
good fit could alternatively be obtained with an optically thin
recombination spectrum, although this would be inconsistent with the
limited spectroscopic observations of other sources across the Balmer
jump suggesting it is not present in emission in those cases. In general, the
temperatures are higher (sometimes much higher) than expected for
quiescent disks, and the UV light sources are also much smaller than
the whole disk. The most likely origin of the stream-impact point can
probably be discounted as the requisite accretion rates for both
GU~Mus and Aql~X-1 are at least an order of magnitude above estimates,
or limits on, the time-averaged accretion rate.  It is therefore more
likely that the UV emission originates from a hot inner region of the
disc.

\acknowledgments

This work includes observations with the NASA/ESA {\it Hubble Space
  Telescope}, obtained at STScI, which is operated by AURA Inc.\ under
NASA contract No.\ NAS5-26555.  Support for {\it HST} proposal
GO\,10253 was provided by NASA through a grant from STScI.  R.I.H.
also acknowledges support by the National Science Foundation under
Grant No.\ AST-0908789.  This work has made use of the NASA
Astrophysics Data System Abstract Service.

{\it Facilities:} \facility{HST (ACS)}.

\begin{table*}
\caption{Log of ACS/HRC observations. }
\label{LogTable}
\begin{tabular}{lllccrr}
\hline
\noalign{\bigskip}
Object & Date   & Filter & UT range & Total    & \multicolumn{1}{c}{Count}           & \multicolumn{1}{c}{$F_{\lambda}$}                                    \\
       &        &        &          & time (s) & \multicolumn{1}{c}{rate (s$^{-1}$)} & \multicolumn{1}{c}{($10^{-18}$\,erg\,cm$^{-2}$\,s$^{-1}$\,\AA$^{-1}$)} \\
\noalign{\smallskip}
\hline
\noalign{\smallskip}
GU~Mus   & 2004 Oct 12 & F330W & 07:19--09:35 & 1920 & $6.63\pm0.10$  & $14.84\pm0.22$ \rule{8mm}{0mm}\\
         &             & F250W & 07:30--09:11 & 1860 & $2.73\pm0.07$  & $13.07\pm0.32$ \rule{8mm}{0mm}\\
         &             & F220W & 07:47--09:25 & 1280 & $0.75\pm0.08$  & $6.11\pm0.63$ \rule{8mm}{0mm}\\
\noalign{\smallskip}
V518~Per & 2005 Oct 31 & F330W & 12:34--16:30 & 2880 & $0.62\pm0.06$  & $1.39\pm0.12$ \rule{8mm}{0mm}\\
         &             & F250W & 12:44--16:08 & 2440 & $0.10\pm0.04$  & $0.48\pm0.21$ \rule{8mm}{0mm}\\
         &             & F220W & 12:59--16:20 & 1480 & $-0.06\pm0.07$ & $-0.53\pm0.57$ \rule{8mm}{0mm}\\
\noalign{\smallskip}
MM~Vel  & 2004 Dec 11  & F330W & 10:45--13:05 & 1720 & $0.99\pm0.08$  & $2.20\pm0.17$ \rule{8mm}{0mm}\\ 
        &              & F250W & 10:54--12:42 & 1780 & $0.51\pm0.05$  & $2.43\pm0.25$ \rule{8mm}{0mm}\\
        &              & F220W & 11:12--12:54 & 1200 & $-0.01\pm0.07$ & $-0.10\pm0.58$ \rule{8mm}{0mm}\\
\noalign{\smallskip}
Aql~X-1 & 2004 Sep 22  & F330W & 20:10--00:04 & 2880 & $2.18\pm0.06$ & $4.87\pm0.14$ \rule{8mm}{0mm}\\
        &              & F250W & 20:20--23:43 & 2360 & $0.56\pm0.05$ & $2.68\pm0.25$ \rule{8mm}{0mm}\\
        &              & F220W & 20:34--23:54 & 1540 & $0.06\pm0.07$ & $0.49\pm0.60$ \rule{8mm}{0mm}\\
\noalign{\smallskip}
\noalign{\smallskip}
\hline
\end{tabular}\\
\end{table*}

\end{document}